\documentclass[onecolumn,12pt,doublespace]{article}

\usepackage{multirow}
\usepackage[pdftex]{graphicx}
\usepackage{amsfonts,amsmath,amssymb,amsthm,epsfig,subfigure}
\usepackage{epsfig}

\newcommand{\be}{\begin{eqnarray}}
\newcommand{\ee}{\end{eqnarray}}

\def\eg{{\em e.g.},~}
\def\ie{{\em i.e.},~}
\def\cf{{\em cf.},~}

\newcommand{\ben}{\begin{enumerate}}
\newcommand{\een}{\end{enumerate}}
\newcommand{\beq}{\begin{equation}}
\newcommand{\eeq}{\end{equation}}
\newcommand{\beqa}{\begin{eqnarray*}}
\newcommand{\eeqa}{\end{eqnarray*}}
\newcommand{\bit}{\begin{itemize}}
\newcommand{\eit}{\end{itemize}}
\newcommand{\bt}{\begin{tabular}{c}}
\newcommand{\btt}{\begin{tabular}}
\newcommand{\et}{\end{tabular}}

\begin{document}
\title{Roaming charges for customers of\\ cellular-wireless
entrant and incumbent providers\thanks{This
research was supported by NSF CNS grant 1116626.  Contact the
corresponding
author G. Kesidis at  kesidis@gmail.edu}}

\author{
\begin{tabular}{cc}
G. Kesidis, D. Mercer, C. Griffin & S. Fdida\\
The Pennsylvania State University & UPMC\\
 University Park, PA, 16803, USA & Paris, France
\end{tabular}
}

\maketitle

\section{Introduction}

In some countries, 
cellular wireless roaming charges are prohibitive possibly because 
the incumbent access-provider
is actually or essentially state-run, unlike potential entrants,
\ie the regulator (state) is in a conflict of interest.
In other countries, 
\eg France, 
there is a lot of competition
among Internet Service (access) Providers (ISPs). Recently, Free purchased a
spectrum license to compete in the 4G market. Free is an
established discount broadband (wired) ISP likely intending
to bundle its existing offerings with cellular wireless.
The cellular-wireless incumbents such as Orange disputed Free's position
on roaming charges for its customers while Free builds out
its wireless infrastructure and offers highly discounted access
rates to attract customers
(though Free's service is quota limited and considered of poorer quality and
support, the latter particularly through physical store-fronts) 
\cite{OrangeFree}. Orange does lease some of its existing infrastructure
to third-party discount providers such as Virgin Wireless (which
does not offer bundled services in direct competition with Orange).

The Canadian government also recently considered regulating 
roaming charges \cite{CBC13} for similar reasons: the entrant
wishes relatively low roaming charges so as to be able to offer
competitive prices and attract customers while not operating 
at a severe loss in the short term, whereas the incumbents 
demand that their higher operating costs are respected including
minimally profitable legacy services (\eg telephony) that they
are obliged to maintain. Left to the incumbents, roaming charges
may rise to create a barrier to entry into the cellular wireless
market.

We consider two competing cellular wireless access providers,
indexed 1 and 2, that serve overlapping areas.
The coverage lapses of the entrant 2 can be accommodated by 
the 
incumbent 1, but not vice versa, \ie
    the entrant 2 has much less deployed infrastructure than incumbent 1.
So as not to trivialize matters, we assume that the entrant
attempts to maintain profitability while it grows its
cellular wireless access infrastructure.
The roaming charge is assumed regulated, \ie it's not
controlled by either access provider.



\section{Asymmetric case with 
large incumbent (1) and
a small entrant (2)}

\subsection{Demand response and ISP-player utilities}

Let $\phi$ be the fraction of the entrant's demand that roams and
let $r$ be the associated roaming charges per unit demand.
We assume that 
the effective price of the entrant's customers is
$p_2 +  \phi r$,
where $p_k$ is the access price for the $k$th ISP. 

We use a model of demand that considers
both response to price and 
congestion as in \eg \cite{Johari10}, but in our
model the congestion based term {\em implicitly} depends
on demand itself so that 
the incumbent and entrant demands, respectively, satisfy
\cite{EconCom13-corrected}: 
\be
D_1  & = &D_{\max}(1-\delta \bar{p})
\frac{p_2+\phi r}{p_1+p_2+\phi r}
g_1(D_1+\phi D_2,B_1)
\label{D1}\\
D_2  & = & D_{\max}(1-\delta \bar{p})
\frac{p_1}{p_1+p_2+\phi r}
g_2((1-\phi)D_2,B_2)
\label{D2}
\ee
where:
\begin{itemize}
\item The first term accounts for how {\em total} demand
is sensitive to price, here assumed 
linearly decreasing with average\footnote{More 
general or complex forms
of demand response could be numerically considered, including
one instead involving a ``social" average price implicitly dependent
on demand, \eg
$\bar{p}  = (D_1 p_1 + D_2 (p_2+\phi r))/(D_1+D_2)$.}
access price from maximum,
\be\label{pavg}
\bar{p} & = & \frac{p_1 + (p_2+\phi r)}{2},
\ee
where $\delta D_{\max}$ is the demand sensitivity to price.
\item The second (competition) factor models how demand is divided between
the ISPs based on their access price, \ie ISP $k$'s demand is inversely 
proportional to $p_k$, \eg \cite{ReArch10}.
\item  The third
factor models how demand depends on congestion 
via a ``demand capacity" parameter $B$ and ``headroom"
parameter $\gamma>0$
\cite{EconCom13-corrected}: $g(D,B)$
decreases with $D<B-\gamma$ and increases with $B$, 
$g(0,B)\equiv 1\equiv g(D,\infty)$ and
$g(B-\gamma,B)\equiv 0$.
For example, a nonlinear $g(\cdot,B)$ 
based on the
mean delay of an M/M/1 queue
\cite{Wolff89} is
$g(D,B)=(1-\gamma/(B-D))/(1-\gamma/B)$.
A simpler, linearized congestion factor is $g(D,B) = 1-D/(B-\gamma)$.
\end{itemize}

The incumbent and entrant ISP utilities are, respectively,
\beqa
U_1 (p_1,p_2) & = & (p_1-c_{d,1}) D_1 +  (r-c_{d,1})\phi D_2 - c_{b,1} B_1\\
U_2 (p_2,p_1) & = & (p_2-(1-\phi)c_{d,2}) D_2  - c_{b,2} B_2
\eeqa
where $c_b$ is demand-independent 
operational expenditures (op-ex), including amortized
capital expenditures, cap-ex) per-unit infrastructure resource ($B$),
and $c_d$ is per-unit demand-dependent op-ex.

\subsection{Game set-up, simplifications and discussion of objectives}

We now assume that demand-dependent op-ex
$c_d$ are negligible
for analytical simplicity.  The infrastructure based costs will not impact
Nash equilibrium prices and will complicate our notion of ``fairness" 
regarding roaming charges, \cf (\ref{fairness-cond0}).
So in this section, we will assume $c_b\approx 0$ too.
We also consider the system free of congestion\footnote{Without 
congestion and under the model of received demand
inversely proportionate to price, the average
``social" price (footnote 1) is the harmonic average,
$\bar{p}^{-1} = (p_1^{-1} + p_2^{-1})/2$.}, 
\ie $g_1,g_2\approx 1$ in (\ref{D1}) and (\ref{D2}).
Thus, we will consider the following utility functions
\be
U_1 (p_1,p_2)/D_{\max} & = &  p_1
(1-\delta \bar{p})\frac{p_2+\phi r}{p_1+p_2+\phi r} 
 + r\phi
(1-\delta \bar{p}) \frac{p_1}{p_1+p_2+\phi r} \label{U1-simple}\\
U_2 (p_2,p_1)/D_{\max} & = & p_2
(1-\delta \bar{p}) \frac{p_1}{p_1+p_2+\phi r} \label{U2-simple}
\ee

Assuming both sets of customers roam in the same domain 
(that of the incumbent), we can take the roaming factor 
\be\label{phi}
\phi & = & \frac{B_1-B_2}{B_1} ~=~ 1-\frac{B_2}{B_1}.
\ee

Thus, the simplified system has three positive parameters
in addition to initial prices (play-actions): $\phi,\delta,r$.
Again, we assume that the roaming factor $r$ is set by a regulator.

\subsection{Objectives}


Assuming $B_1\gg B_2$ ($\phi\approx 1$),
our objective herein is to determine the
Nash equilibrium (NE) prices $p^*_1,p^*_2$ and see 
how the NE utilities $U^*$ depend on the roaming charge,
$r$. 
The impact of the demand-sensitivity-to-price parameter
$\delta D_{\max}$ will be to simply shift and scale
the NE prices.

Specifically, we want to see whether 
{\em fairness} is achieved at NE, \ie
whether net revenue is proportional to expenditures:
\be\label{fairness-cond0}
\frac{U_1^*(r)}{B_1} & = & \frac{U_2^*(r)}{B_2},
\ee
equivalently under (\ref{phi}), whether 
\be\label{fairness-cond1}
(1-\phi)U_1^*(r)-U_2^*(r) & = & 0.
\ee

\subsection{Analytical results for simplified system}

An ``interior" (strictly positive, finite) 
solution to the first-order necessary conditions (FONC),
\be\label{FONC}
\frac{\partial U_1}{\partial p_1} & =~ 0 ~= & 
\frac{\partial U_2}{\partial p_2},
\ee
of (\ref{U1-simple}) and (\ref{U2-simple})
is a symmetric one where {\em both} NE prices
\be
p_1^*,p_2^* & = & 
\frac{1}{4\delta}\left(
-2\delta r \phi +1+ \sqrt{4\delta r\phi+1} 
\right) ~>~0 \label{NE}
\ee
when
\be\label{interior-cond}
\delta r \phi & < & 2,
\ee
\ie when there are feasible prices $(p_1,p_2)$
according to 
\beqa
p_1+p_2 & \leq & - r\phi + 2/\delta.
\eeqa
Indeed, it can be directly verified that
$p_1^*+p_2^* = 2p_1^*=2p_2^* < -r\phi + 2/\delta$.

These prices also satisfy 
$\partial^2 U_1/\partial p_1^2,~
\partial^2 U_2/\partial p_2^2 < 0$, with
strictly positive utilities $U_1^*,U_2^*>0$ under
(\ref{interior-cond}), so that 
indeed they are ``locally  Nash."
The other solutions of the FONC 
(\ref{FONC}) either have $p_2^*<0$
(\ie extraneous)
or $p_1^*=0$.
But  if $p_1^*=0$ then $U_1^*=0=U_2^*$.
Also, if $p_2=0$ then $\partial U_2/\partial p_2 >0$.
So, there
are no boundary Nash equilibria and thus
(\ref{NE}) is the unique Nash equilibrium.

It can also be directly shown for this model that 
there is a solution 
\be\label{rstar}
r^* & := & \frac{2(2-\phi)}{\delta(4-3\phi)^2}
\ee
at which (\ref{fairness-cond1}) holds. 
Note that $r^* < 2(\delta\phi)^{-1}$ for $0\leq \phi <1$.
Roaming prices $r<r^*$ favor the entrant, otherwise
the incumbent, \cf next subsection.

\subsection{Numerical results}

\begin{figure}[ht]
\begin{center}
\includegraphics[width=3.25in]{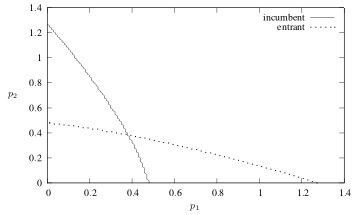}
\caption{Best-response curves
for $\phi=0.9,\delta=1,r=0.8$}\label{best-response-fig}
\end{center}
\end{figure}

We considered the example with $\delta=1$ and $\phi=0.9=1-B_2/B_1$.
For the model without congestion under (\ref{interior-cond}),
the two ``best response" curves for $r=0.8$ are given
in Figure \ref{best-response-fig}.
That is, the first curve has vertical
distance to the $x$-axis,
\beqa
p_2^*(p_1) & = & \mbox{arg}\max_{0\leq p_2\leq p_{\max}}
U_2(p_2,p_1).
\eeqa
The second curve has horizontal
distance to the $y$-axis,
\beqa
p_1^*(p_2) & = & \mbox{arg}\max_{0\leq p_1\leq p_{\max}}
U_1(p_1,p_2).
\eeqa
These curves meet at the Nash equilibrium, here 
$(p_1^*,p_2^*)\approx (0.38,0.38)$ which is consistent with 
(\ref{NE}).

\begin{figure}[ht]
\begin{center}
\includegraphics[width=3.25in]{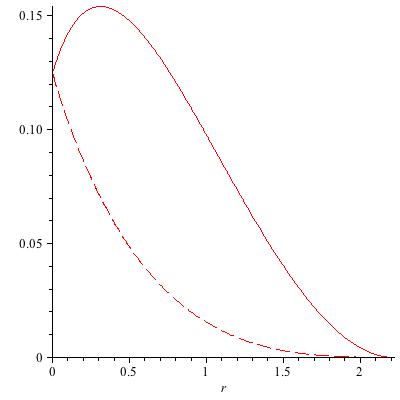}
\caption{Utilities of the incumbent
$U_1^*(r))/D_{\max}$  (solid line) and entrant
$U_2^*(r)/D_{\max}$ (dashed  line) 
versus roaming charge $r$ with $\phi=0.9,\delta=1$}\label{utilities-fig}
\end{center}
\end{figure}

We also numerically verified that
Nash-equilibrium utilities $U_1^* (r)$ and $U_2^*(r)$ are 
positive, see Figure \ref{utilities-fig}.   Both
decrease and reach zero at $r=2(\delta\phi)^{-1}=20/9$ (again, 
a point where the only
feasible prices are $p_1=0=p_2$). Note that utilities are
generally higher for lower roaming charges in our simple model.

\begin{figure}[ht]
\begin{center}
\includegraphics[width=3.25in]{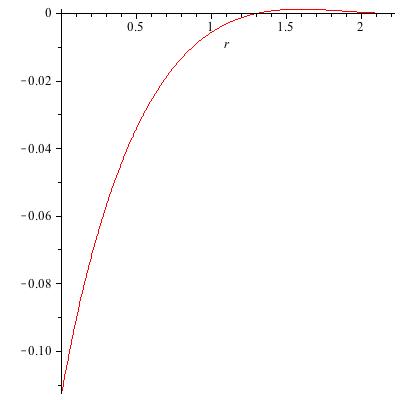}
\caption{$((1-\phi)U_1^*(r)-U_2^*(r))/D_{\max}$ 
versus roaming charge 
$r$ with $\phi=0.9,\delta=1$}\label{fairness-fig}
\end{center}
\end{figure}

In Figure \ref{fairness-fig}, we plot the 
``fairness" expression in 
(\ref{fairness-cond1}) and verify that 
(\ref{fairness-cond1}) holds at 
$r^* \approx 1.3 < 20/9$, consistent with  (\ref{rstar}).
And we see from this figure how roaming prices
$r<r^*$ favor the entrant, otherwise the incumbent.


\section{Future work}

Recall that we assumed negligible op-ex, \ie assumed 
$c_b,c_d\approx 0$. Depending on the situation,
op-ex
per unit demand might be lower for the  entrant 
(\eg only present in areas involving cheaper deployment costs
and higher customer density) or
for the incumbent (generally owing to greater scale of operations).
Such a discrepancy 
could be accounted for in our ``fairness"
condition (\ref{fairness-cond0}).
Future work will also consider
the effects of congestion, 
more complex price-competition and price-sensitivity models,
and multiple competing incumbents and entrants
(as in \cite{ReArch10}).

\subsection*{Acknowledgment:}
We wish to thank Dr. Ashraf Al Daoud for a preliminary discussion
on regulating roaming charges.

\bibliographystyle{plain}

\end{document}